\newcommand{\htofourb}{\ensuremath{H\rightarrow aa\rightarrow  b\overline{b}b\overline{b}}\xspace}
\newcommand{\ctaua}{\ensuremath{c\tau_a}\xspace}
\newcommand{\ma}{\ensuremath{m_a}\xspace}
 
\newcommand{\drmax}{\ensuremath{\Delta R_{\mathrm{max}}}\xspace}
\newcommand{\ntrk}{\ensuremath{n_{\text{trk}}}\xspace}
\newcommand{\dv}{\ensuremath{\mathrm{DV}}\xspace}

\documentclass[cernpreprint, atlasdraft=false, texlive=2023, UKenglish, texmf]{atlasdoc}
 
\usepackage{atlaspackage}
\usepackage{pdflscape}
\usepackage{atlasbiblatex}

\usepackage{atlasphysics}
 
\usepackage{PMGRefs-defs}
 
\usepackage{multirow}
 
\graphicspath{{logos/}{figures/}}

\PreprintIdNumber{CERN-EP-2021-106}
 
\AtlasTitle{Search for exotic decays of the Higgs boson into long-lived particles
in $pp$ collisions at $\sqrt{s} = 13$~TeV using displaced vertices in the ATLAS inner detector}
 
\AtlasDOI{10.1007/JHEP11(2021)229}
\AtlasJournalRef{JHEP 11 (2021) 229}
 
\AtlasAbstract{
A novel search for exotic decays of the Higgs boson into pairs of long-lived
neutral particles, each decaying into a bottom quark pair, is performed
using 139~fb$^{-1}$ of $\sqrt{s} = 13$~\TeV\ proton--proton collision data
collected with the ATLAS detector at the LHC.
Events consistent with the production of a Higgs boson in association with a
leptonically decaying $Z$ boson are analysed.
Long-lived particle (LLP) decays are reconstructed from inner-detector tracks
as displaced vertices with high mass and track multiplicity relative to Standard
Model processes.
The analysis selection requires the presence of at least two displaced vertices,
effectively suppressing Standard Model backgrounds.
The residual background contribution is estimated using a data-driven technique.
No excess over Standard Model predictions is observed, and upper limits
are set on the branching ratio of the Higgs boson to LLPs.
Branching ratios above $10\%$ are excluded at $95\%$ confidence level for LLP mean proper lifetimes
$c\tau$ as small as 4~mm and as large as 100~mm.
For LLP masses below 40~\GeV, these results represent the most stringent
constraint in this lifetime regime.
}
 
\author{The ATLAS Collaboration}
 
\AtlasRefCode{EXOT-2018-57}

\hypersetup{pdftitle={ATLAS document},pdfauthor={The ATLAS Collaboration}}
 
\begin{document}
 
\maketitle
 
\tableofcontents
 
\section{Introduction}
\label{sec:intro}

The discovery of a new particle consistent with the Higgs boson at the Large
Hadron Collider in 2012 revealed the final missing piece
of the Standard Model~(SM) of particle physics~\cite{HIGG-2012-27,CMS-HIG-12-028}.
Since then, a physics programme has emerged which aims to scrutinize the
properties of the Higgs boson and the nature of the Higgs sector.
Assuming that new physics does not strengthen the couplings of the Higgs boson to $W/Z$ bosons, current experimental constraints allow for branching ratios of the Higgs
boson to visible beyond-the-SM (BSM)
states of up to 21\%~\cite{HIGG-2018-57,CMS-HIG-17-031}.
Further, many theories that aim to solve the shortcomings of the Standard
Model point to the Higgs boson as a possible portal to new
physics~\cite{Patt:2006fw}, with exotic Higgs boson decay modes being the primary
phenomenological consequence and means of discovery~\cite{Curtin:2015fna,Bock_2010,Strassler_2007}.
 
Decays of the Higgs boson into particles with macroscopic mean proper lifetimes ($c\tau\gtrsim100~\mu\mathrm{m}$), known as
long-lived particles (LLPs), are predicted by top-down theories and are
further motivated by bottom-up considerations.
Examples of top-down theories include those based on the concept of Neutral
Naturalness, such as Folded SUSY~\cite{Burdman_2007,Cohen_2016} and Quirky
Little Higgs~\cite{Cai_2009}.
These models aim to solve the hierarchy problem by introducing quark partners
that are not charged under the SM $\mathrm{SU}(3)_\mathrm{C}$ group, but instead under some new
SU(3)$^{\prime}$ gauge group~\cite{Curtin:2015fna}.
While conventional solutions to the hierarchy problem with coloured quark partners, such as supersymmetry~\cite{Golfand:1971iw,Volkov:1973ix,Wess:1974tw,Wess:1974jb,Ferrara:1974pu,Salam:1974ig}, are becoming increasingly constrained experimentally, the production cross section for such uncoloured quark partners is greatly reduced, easily evading  experimental constraints while remaining discoverable at the LHC.
If the quark partners are sufficiently heavy,
pure-SU(3)$^{\prime}$ glueballs (SU(3)$^{\prime}$-neutral bound states composed of SU(3)$^{\prime}$ mediators) would lie at the bottom of the BSM particle mass spectrum.
Glueball masses in the range 10--60~\GeV\ are preferred theoretically, and could be produced in decays of the Higgs boson.  At the bottom of the BSM mass spectrum, the only decay mode available to such glueballs would be to SM fermions through an off-shell Higgs boson (with Yukawa-ordered branching ratios), causing decays into $b\bar{b}$ to dominate with lifetimes anywhere in the micrometer to kilometer range~\cite{Curtin:2015fna}.
 
Bottom-up scenarios, such as hidden valleys~\cite{HV_DV,EchoesHV} and minimal
scalar extensions to the SM, are also well motivated because the SM Higgs field
admits the renormalizable portal interaction $\epsilon |S|^2 H^{\dagger} H$.
These simplified scenarios often arise naturally in the infrared limit
of more complete theories such as Folded SUSY~\cite{Burdman_2007}, and provide
generic examples of scenarios in which the Higgs boson decays into LLPs.
 
\begin{figure}
\centering
\includegraphics[width=0.6\textwidth]{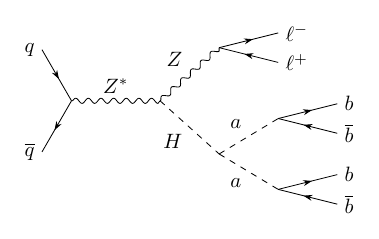}
\caption{Tree-level Feynman diagram representing $ZH$ production with
$H\rightarrow aa\rightarrow  b\bar{b}b\bar{b}$.}
\label{fig:feyn}
\end{figure}

This paper considers a simplified model in which a new electrically neutral pseudoscalar $a$
boson is produced in pairs through decays of the Higgs boson, with the $a$
boson subsequently decaying exclusively into $b\bar{b}$, as shown in Figure~\ref{fig:feyn}.
The mean proper lifetime \ctaua and mass \ma of the $a$~boson are treated as free parameters.  Aside from the $Haa$ interaction, the Higgs boson is assumed to be that of the SM.
 
Several searches for $\htofourb$ decays have been performed at the LHC,
optimized for different regimes of \ctaua.
Searches based on standard $b$-tagging techniques were performed by ATLAS, with
sensitivity for $\ctaua\lesssim1 $~mm~\cite{HIGG-2017-05,HDBS-2018-47}.
While these searches focused on the topology of associated $VH$ ($V = W,Z$)
production of the Higgs boson, where leptonic $V$ decays provide a trigger
object, other searches have targeted gluon--gluon fusion production of the Higgs
boson, which has a much larger cross section.  For longer $a$~boson lifetimes, standard $b$-tagging algorithms become inefficient compared to specialized reconstruction algorithms.
The LHCb experiment performed a search for displaced vertices (DVs) resulting
from LLP decays, obtaining sensitivity in the range
1~mm~$\lesssim\ctaua\lesssim$~0.1~m~\cite{Aaij:2265567}.
The CMS experiment also performed a DV-based search, yielding sensitivity in
the range 1~mm~$\lesssim\ctaua\lesssim$~1~m for
$\ma\geq40$~\GeV~\cite{Sirunyan:2746387}.
To identify LLP decays in the hadronic calorimeter, the ATLAS experiment
selects jets that deposited anomalously more energy in the hadronic calorimeter than in the electromagnetic calorimeter
(both at trigger level~\cite{TRIG-2012-02} and offline), providing sensitivity for
0.1~m~$\lesssim\ctaua\lesssim$~10~m~\cite{EXOT-2017-25}.
Finally, ATLAS extends this sensitivity to $\ctaua\lesssim$~100~m with
searches for LLPs decaying within the muon spectrometer, by requiring a cluster of muon regions-of-interest from the first-level trigger~\cite{TRIG-2012-02} and a DV reconstructed from muon spectrometer tracklets~\cite{EXOT-2017-05,EXOT-2018-61}.
 
The search presented here targets LLP decays within the ATLAS inner-detector
volume, resulting in the presence of at least two DVs that may be reconstructed
from inner-detector tracks.
The specialized algorithms for reconstruction of LLP decays are computationally
expensive and therefore cannot be used at trigger level or in the standard
ATLAS reconstruction chain~\cite{SUSY-2016-08}.
As a result, the $ZH$ associated Higgs boson production topology is probed in
this analysis, where $Z \rightarrow \ell\ell$ ($\ell = e,\mu$) decays provide
highly efficient trigger objects.
Based on the results of the standard event reconstruction, a small subset of the data is selected (as described in Section~\ref{sec:select}) for reprocessing with dedicated tracking and vertexing algorithms that can reconstruct displaced LLP decays~\cite{ATL-PHYS-PUB-2017-014,ATL-PHYS-PUB-2019-013}.
Finally, events are required to contain at least two DVs with high mass and
track multiplicity relative to SM processes, effectively suppressing
backgrounds.
The residual background is estimated using a data-driven method and validated
using an orthogonal signal-free data sample.
 
This is the first search for Higgs boson decays into LLPs to make use of this analysis
methodology.  Although it is assumed that the $a$~boson is neutral and decays exclusively into $b\bar{b}$, these signal features are not explicitly exploited, so sensitivity is maintained for new physics scenarios with charged intermediate states and/or other hadronic decay modes.  With this strategy, sensitivity to LLPs with 2~mm~$\lesssim\ctaua\lesssim$~0.2~m is obtained.


\section{ATLAS detector}
\label{sec:detector}

\newcommand{\AtlasCoordFootnote}{
ATLAS uses a right-handed coordinate system with its origin at the nominal interaction point (IP)
in the centre of the detector and the \(z\)-axis along the beam pipe.
The \(x\)-axis points from the IP to the centre of the LHC ring,
and the \(y\)-axis points upwards.
Cylindrical coordinates \((r,\phi)\) are used in the transverse plane,
\(\phi\) being the azimuthal angle around the \(z\)-axis.
The pseudorapidity is defined in terms of the polar angle \(\theta\) as \(\eta = -\ln \tan(\theta/2)\).
Angular distance is measured in units of \(\Delta R \equiv \sqrt{(\Delta\eta)^{2} + (\Delta\phi)^{2}}\).}
 
The ATLAS detector~\cite{PERF-2007-01} at the LHC covers nearly the entire solid angle around the collision point.\footnote{\AtlasCoordFootnote}
It consists of an inner tracking detector surrounded by a thin superconducting solenoid, electromagnetic and hadron calorimeters,
and a muon spectrometer incorporating three large superconducting air-core toroidal magnets.
 
The inner-detector system (ID) is immersed in a \SI{2}{\tesla} axial magnetic field
and provides charged-particle tracking in the range \(|\eta| < 2.5\).
The high-granularity silicon pixel detector covers the vertex region and typically provides four measurements per track,
the first hit normally being in the insertable B-layer installed before Run~2~\cite{ATLAS-TDR-19,PIX-2018-001}.  The geometry of the pixel detector consists of concentric barrel layers with $r= 33$, 50.5, 88.5, and 122.5~mm in the central region, and three disks in each of the endcaps at $|z| = 495$, 580, and 650~mm.
It is followed by the silicon microstrip tracker (SCT), consisting of barrel layers at $r=299$, 371, 443, and 514~mm, spanning $|z| < 746$~mm, and nine wheels in each of the endcaps with $854 < |z| < 2720$~mm.  The SCT usually provides eight measurements per track.
These silicon detectors are complemented by the transition radiation tracker (TRT),
which enables radially extended track reconstruction up to \(|\eta| = 2.0\).
The TRT also provides electron identification information
based on the fraction of hits (typically 30 in total) above a higher energy-deposit threshold corresponding to transition radiation.
 
The calorimeter system covers the pseudorapidity range \(|\eta| < 4.9\).
Within the region \(|\eta|< 3.2\), electromagnetic calorimetry is provided by barrel and
endcap high-granularity lead/liquid-argon (LAr) calorimeters,
with an additional thin LAr presampler covering \(|\eta| < 1.8\)
to correct for energy loss in material upstream of the calorimeters.
Hadron calorimetry is provided by the steel/scintillator-tile calorimeter,
segmented into three barrel structures within \(|\eta| < 1.7\), and two copper/LAr hadron endcap calorimeters.
The solid angle coverage is completed with forward copper/LAr and tungsten/LAr calorimeter modules
optimized for electromagnetic and hadronic energy measurements respectively.
 
The muon spectrometer (MS) comprises high-precision tracking chambers measuring
the deflection of muons in a magnetic field generated by the superconducting
air-core toroidal magnets.
The field integral of the toroids ranges between \num{2.0} and \SI{6.0}{\tesla\metre}
across most of the detector.
A set of precision chambers covers the region \(|\eta| < 2.7\) with three layers of monitored drift tubes,
complemented by cathode-strip chambers in the forward region, where the background is highest.
The muon trigger system covers the range \(|\eta| < 2.4\) with resistive-plate chambers in the barrel, and thin-gap chambers in the endcap regions.
 
Interesting events are selected by the first-level trigger system implemented in custom
hardware using information from the muon and calorimeter systems,
followed by selections made by algorithms implemented in software in the high-level
trigger~\cite{TRIG-2016-01}.
The first-level trigger accepts events from the \SI{40}{\MHz} bunch crossings at a rate below \SI{100}{\kHz},
which the high-level trigger further reduces in order to record events to disk at about \SI{1}{\kHz}.
 
An extensive software suite~\cite{ATL-SOFT-PUB-2021-001} is used for real and simulated data reconstruction
and analysis, for operation and in the trigger and data acquisition systems of the experiment.


\section{Data and simulated events}
\label{sec:data}

This analysis uses $139.0~\mathrm{fb}^{-1}$ of $\sqrt{s} = 13$~\TeV\ $pp$ collision
data, collected by the ATLAS experiment from 2015 to 2018.
Only data collected during stable beam conditions in which all detector
subsystems were operational are considered~\cite{DAPR-2018-01}.
Events were selected using the lowest unprescaled single-lepton ($e$, $\mu$)
triggers available during each data-taking year~\cite{TRIG-2018-05,TRIG-2018-01}.
Additional events selected with single-photon and diphoton triggers are used
for validation of the background modelling strategy.
 
Simulated signal and background events were used to develop the analysis,
optimize the event selection criteria, measure the signal selection efficiency,
and assess sources of systematic uncertainty.
The effect of multiple proton--proton interactions in the same or neighbouring
bunches (pile-up) was modelled by overlaying the hard-scatter process with
simulated inelastic proton--proton scattering events generated with
\PYTHIA[8.186]~\cite{Sjostrand:2014zea} using the NNPDF2.3lo PDF set~\cite{Ball:2012cx} and
the A3 set~\cite{ATL-PHYS-PUB-2016-017} of tuned parameters.
The generated events were then run through a detailed
$\GEANT4$~\cite{Agostinelli:2002hh} simulation of the ATLAS detector to
simulate the detector response~\cite{SOFT-2010-01} and reconstructed using the same algorithms
as used for the data.
 
\subsection{Signal simulation}
 
Samples of simulated events with a Higgs boson produced in association with a leptonically-decaying $Z$ boson
were generated using \POWHEGBOX[v2]~\cite{Nason:2009ai,Alioli:2010xd,Nason:2004rx,Frixione:2007vw}.
The \POWHEGBOX prediction is accurate to next-to-leading order (NLO) for $ZH + \textrm{1-jet}$ production.
Amplitudes for QCD virtual/loop corrections were constructed through the interface to the \gosam\
package~\cite{Cullen:2011ac}.
The loop-induced $gg\to ZH$ process was generated separately at leading order
with \POWHEGBOX.
In all cases, the PDF4LHC15nlo PDF set~\cite{Butterworth:2015oua} was used.
The simulated samples are normalized to SM cross sections calculated at NNLO
in QCD with NLO electroweak corrections for $q\bar{q} \to ZH$ and at NLO
and next-to-leading-logarithm accuracy in QCD for $gg \to
ZH$~\cite{deFlorian:2016spz,Ciccolini:2003jy,Brein:2003wg,Brein:2011vx,Altenkamp:2012sx,Denner:2014cla,Brein:2012ne,Harlander:2014wda}, neglecting effects from potential $Haa$ interactions.

The decay of the Higgs boson into two spin-0 $a$~bosons and the subsequent
decay of each $a$~boson into a pair of $b$-quarks were simulated with \PYTHIA[8.212]~\cite{Sjostrand:2014zea}.
The coupling of the $a$~boson to $b$-quarks is assumed to be that of a
pseudoscalar; however, the analysis is insensitive to the parity of the $a$
boson.
\PYTHIA[8.212] was also used for parton showering and hadronization, as well as underlying-event
simulation (using the AZNLO CTEQ6L1 set of tuned parameters~\cite{STDM-2012-23}).
Masses of the $a$~boson in the range 16--55~\GeV\ were considered, and
statistically independent event samples were produced with $c\tau_a$
of 10~mm, 100~mm, and 1~m for each mass.
These samples were reweighted to obtain samples corresponding to alternative
lifetimes, where the weights were derived from the ratio of lifetime probability
density functions.
 
The impact of using a different parton shower model on the $\pt$ of the $ZH$
system was evaluated by comparing the nominal $ZH, H \rightarrow aa$ samples
with an additional sample produced with the \POWHEGBOX[v2] generator using the
NNPDF3.0nnlo set of PDFs~\cite{Ball:2014uwa} and interfaced with
\HERWIG[7.2.1]~\cite{Bahr:2008pv,Bellm:2020jjp} using the H7.1-Default set of
tuned parameters~\cite{Bellm:2017jjp} and the MMHT2014nnlo PDF
set~\cite{Harland-Lang:2014zoa}.
 
\subsection{Background simulation}
 
While the background modeling strategy used in this analysis is fully data
driven, simulated $Z+\text{jets}$ events are used to optimize the analysis
selections and derive systematic uncertainties.
The production of $Z+\mathrm{jets}$ was simulated with the
\SHERPA[2.2.1]~\cite{Bothmann:2019yzt}
generator using NLO matrix elements for up to two partons, and LO matrix elements
for up to four partons calculated with the Comix~\cite{Gleisberg:2008fv}
and \openloops~\cite{Buccioni:2019sur,Cascioli:2011va,Denner:2016kdg} libraries. They
were matched with the \SHERPA parton shower~\cite{Schumann:2007mg} using the MEPS@NLO
prescription~\cite{Hoeche:2011fd,Hoeche:2012yf,Catani:2001cc,Hoeche:2009rj}
using the set of tuned parameters developed by the \SHERPA authors.
The NNPDF3.0nnlo set of PDFs~\cite{Ball:2014uwa} was used.


\section{Event reconstruction}
\label{sec:objSelect}

 
Tracks are reconstructed from the collection of hits in the ID using a
combinatorial Kalman filter~\cite{FRUHWIRTH1987444}.  The standard ATLAS track
reconstruction algorithm is optimized for reconstruction of tracks
that originate in the vicinity of the IP, and is not
efficient for reconstructing displaced tracks corresponding to charged
particles produced in LLP decays (although it is generally insensitive to the delayed arrival time of the decay products).
To reconstruct these tracks, a secondary \emph{large radius tracking} (LRT)
algorithm is run on a subset of the data, taking as input the hits left over
from the standard tracking procedure~\cite{ATL-PHYS-PUB-2017-014}.
The LRT algorithm has looser
requirements than the standard track reconstruction, particularly on the transverse and longitudinal impact
parameters with respect to the primary vertex ($d_0$ and
$z_0$).
This algorithm is computationally expensive, but improves the overall reconstruction efficiency for hadrons produced at $r=50$~(300)~mm from about 20\%~(5\%) to 90\%~(30\%)\footnote{The LRT reconstruction efficiency has a non-trivial dependence on track kinematics.  The efficiencies quoted here were calculated from a simulated sample of LLP decays, yielding relatively-soft hadrons with \pt on the order of a few GeV~\cite{ATL-PHYS-PUB-2017-014}.}.
Hadrons produced with $r>300$~mm traverse an insufficient number of tracker
layers for reconstruction.
 
Events are required to contain a primary vertex (PV) associated with the $pp$
hard scatter.
The PV must have at least two tracks, each with $p_{\text{T}} > 500$~\MeV.
If more than one PV candidate satisfies these criteria, the candidate with the largest
$\sum p_\text{T}^2$ is selected, where the sum extends over tracks associated
with the vertex.
 
Electron candidates are reconstructed by matching clusters of energy in the
calorimeters to reconstructed tracks~\cite{EGAM-2018-01}.
Candidates are required to be within the fiducial region $|\eta| < 2.47$, and
not within the calorimeter transition region ($1.37 < |\eta| < 1.52$).
To reduce the background from non-prompt electrons and photon conversions,
electrons must pass the Medium likelihood-based identification working
point~\cite{EGAM-2018-01}.
Electron candidates are additionally required to have $|z_0 \sin \theta| < 0.5$~mm
and $|d_0|/\sigma(d_0) < 5$, where $\sigma(d_0)$ is the
uncertainty in the transverse impact parameter.
 
Photon candidates are similarly reconstructed from clusters of energy in the
calorimeters.
Candidates are required to have $|\eta| < 2.47$ and satisfy the Loose
identification criteria~\cite{EGAM-2018-01}.
 
Muon candidates are reconstructed from track segments in the muon spectrometer
and matched to ID tracks~\cite{MUON-2018-03,PERF-2015-10}.
A combined track is then formed using the information from both detector
subsystems.
Candidates are required to pass the Medium reconstruction
working point~\cite{PERF-2015-10} and are further required to have $|\eta| < 2.5$, $|z_0 \sin
\theta| < 0.5$~mm, and $|d_0|/\sigma(d_0) < 3$.
 
Hadronic jets are reconstructed from topological clusters~\cite{PERF-2014-07} of energy
deposits in the calorimeter, using the anti-$k_{\text{t}}$
algorithm~\cite{Cacciari:2008gp} with a radius parameter $R=0.4$.
Jets are calibrated using simulated events according to the procedure described
in
Refs.~\cite{PERF-2011-03,PERF-2012-01,PERF-2014-03,ATLAS-CONF-2015-002,ATL-PHYS-PUB-2015-015},
followed by a residual calibration that takes into account the differences
between data and simulation, and different detector response in the central
($|\eta| < 0.8$) versus forward ($|\eta| > 0.8$) region through the in
situ calibration techniques described in Refs.~\cite{PERF-2016-04,PERF-2015-05}.
The jets considered in this analysis are required to have $\pt > 20$~\GeV\ and
$|\eta| < 2.5$.
 
The DL1 $b$-tagging algorithm quantifies the likelihood that a jet originated from a $b$-flavour quark versus a gluon $g$ or light-flavour ($u, d, s, c$) quark~\cite{FTAG-2018-01,ATL-PHYS-PUB-2017-013}.  This algorithm exploits several features of $b$-jets, such as the lifetime, mass, and decay kinematics of $b$-hadrons, within a deep feed-forward neural network to distinguish between these hypotheses.  Large values of the DL1 discriminant correspond to $b$-like jets.
 
Two additional track-based observables are used to discriminate between jets initiated by displaced partons resulting from LLP decays and those jets initiated by prompt partons.  The first observable~\cite{CMS-EXO-16-003} exploits the fact that jets with
modest displacement from the IP will contain tracks that
are incompatible with all PV candidates.
Each jet is assigned a vertex-fraction $\alpha_i$ to test its compatibility with PV candidate $i$,
\begin{equation*}
\alpha_i = p_{\mathrm{T}}^{\mathrm{trk}\in i} / p_{\mathrm{T}}^\mathrm{trk},
\end{equation*}
where $p_{\mathrm{T}}^{\mathrm{trk}}$ is the transverse component of the momentum of all tracks geometrically matched to the jet (using a shrinking cone algorithm) and $p_{\mathrm{T}}^{\mathrm{trk}\in i}$ is defined similarly, but considering only the subset of tracks matched to primary vertex candidate $i$.
A track is considered
matched to a PV candidate if $|d_0|<0.5$~mm and $|z_0\sin\theta|<0.3$~mm.
A standard QCD jet should have a large value of $\alpha_i$ for the candidate PV at
which it was produced, while jets originating from the decay of an LLP
should have small values of $\alpha_i$ for all PV candidates.
Therefore, the maximum $\alpha_i$ value for all PVs,
$\alpha_{\mathrm{max}}\equiv\mathrm{max}\left(\alpha_i\right)$, is used to
discriminate between signal and QCD jets, as well as to reject jets originating from
pile-up interactions.
 
The second observable exploits the fact that jets which are more significantly displaced
deposit energy in the calorimeters but contain few reconstructed tracks.
Therefore, the \textit{charged hadron fraction} ($f_\mathrm{ch}$) is defined as
\begin{equation*}
f_\mathrm{ch} = p_{\mathrm{T}}^{\mathrm{trk},\mathrm{prompt}} / \pt^{\mathrm{jet}},
\end{equation*}
where $p_{\mathrm{T}}^{\mathrm{trk},\mathrm{prompt}}$ is the transverse component of the momentum of all tracks geometrically matched to the jet with $|d_0|<0.5$~mm and $\pt^{\mathrm{jet}}$ is the jet transverse momentum.
 
Jets are considered ``displaced'' if they satisfy $\alpha_{\text{max}} < 0.05$ or $f_\text{ch} < 0.045$.  This selection was optimized to provide high signal efficiency for a wide range of LLP proper lifetimes while maximizing background rejection.
 
\subsection{Displaced vertices}
\label{sec:sel_vtx}
 
Displaced vertices are reconstructed from the combined collection of standard
and LRT tracks~\cite{ATL-PHYS-PUB-2019-013}.
The algorithm proceeds in several steps.
First, vertex reconstruction is seeded by pairs of tracks which were
preselected with a series of quality criteria, including hit multiplicity requirements and $p_{\text{T}} > 1$~\GeV.
Further, at least one track must have $|d_0| > 2$~mm.
The compatibility of each possible pair of preselected tracks is then assessed
using the impact parameters of the tracks with respect to the estimated
secondary vertex position, and those deemed loosely compatible are retained.
The tracks are additionally required to pass a hit-pattern requirement which ensures
that the hits associated with each track are compatible with a particle
originating from the position of the seed vertex.
These two-track seed vertices are then combined to form multi-track vertices.
Nearby vertices are then merged, and tracks, including those that failed the
preselection, are ultimately attached to compatible vertices provided that
$\chi^2/n_{\text{dof}}$ of the updated vertex fit is less than 20.
The efficiency of this vertex reconstruction is similar for light-flavour and
heavy-flavour decays of the LLP, once its strong dependence on the multiplicity
of charged LLP decay products is taken into
account~\cite{ATL-PHYS-PUB-2019-013}.
 
Reconstructed DVs are then pruned of tracks which are loosely associated with the vertex;
only those tracks with a transverse (longitudinal) impact parameter with respect
to the DV position satisfying $|d_0^{\dv}|<0.8$~mm ($|z_0^{\dv}|<1.2$~mm)
with an uncertainty $\sigma(d^{\dv}_{0})<0.1$~mm ($\sigma(z^{\dv}_{0})<0.2$~mm) are retained.
After track pruning, vertex properties
are recalculated.
 
A vertex preselection is applied to reject low-quality fake DVs and those
resulting from interactions between high-momentum hadrons and detector material.
These interactions can produce a collimated spray of secondary hadrons,
mimicking an LLP decay.
Displaced vertices are required to fall within the geometric acceptance of the
tracker, defined as $r< 300$~mm and $|z| < 300$~mm.
DVs are further required to have $\chi^2/n_{\text{DoF}} < 5$.
Finally, a \textit{material veto} is applied to reject DVs whose
position is compatible with the location of known detector elements~\cite{SUSY-2016-08}.
 
\begin{figure}
\centering
\subfigure[]{\includegraphics[width=0.49\textwidth]{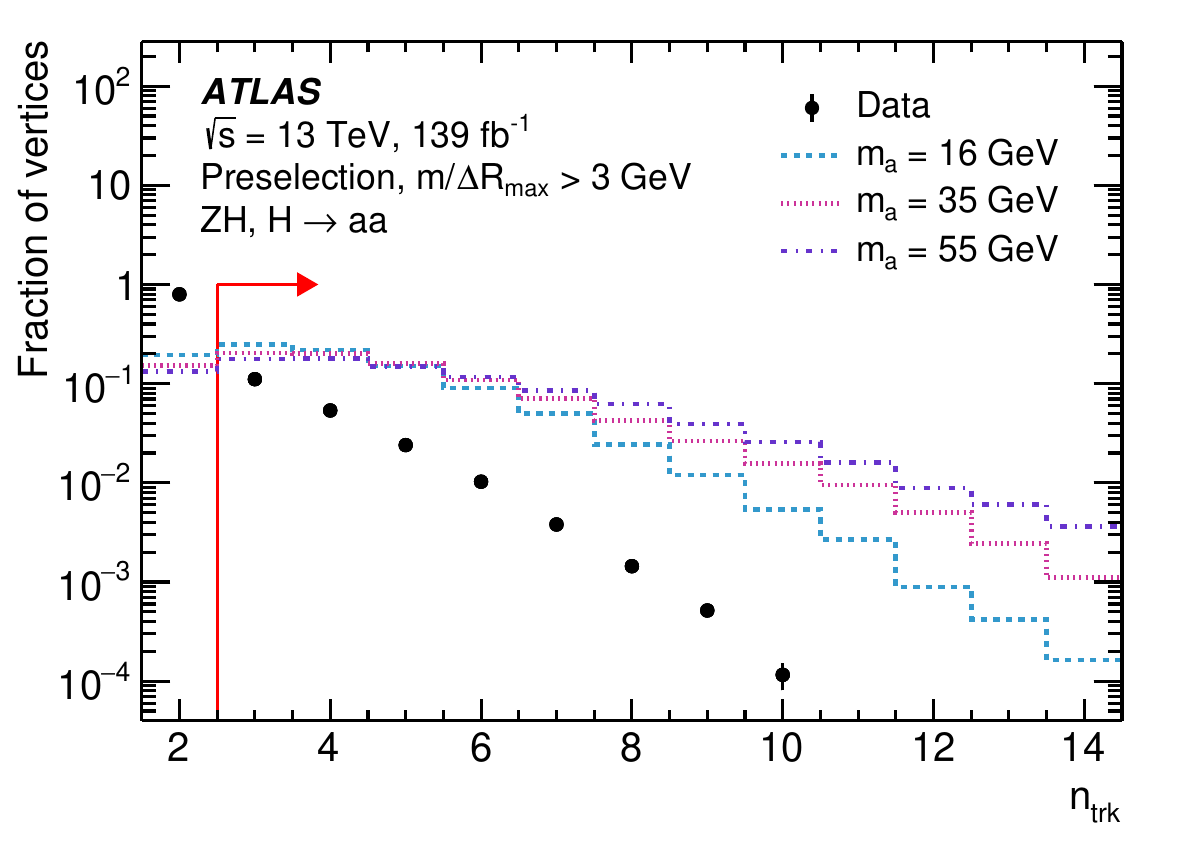}\label{fig:ntrk}}
\subfigure[]{\includegraphics[width=0.49\textwidth]{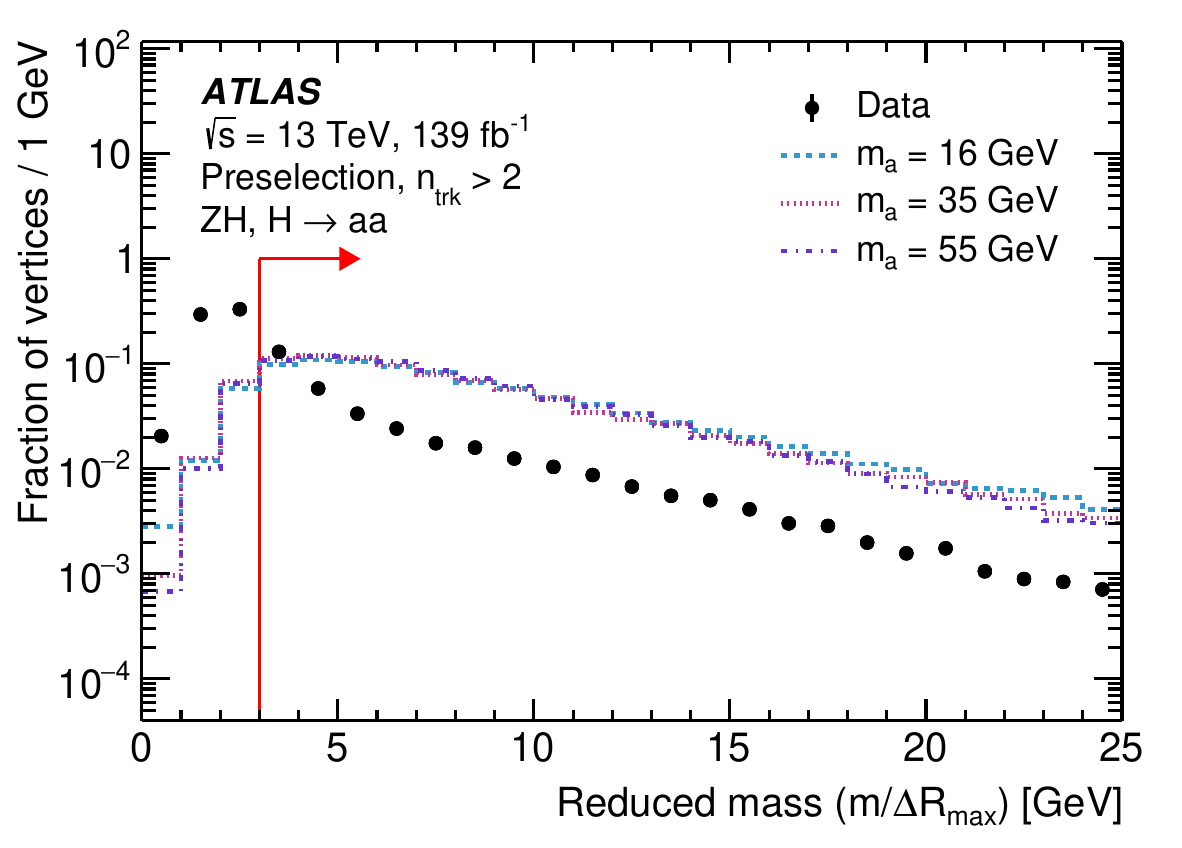}\label{fig:mass}}
\caption{The distribution of (a) \ntrk for preselected DVs with reduced mass exceeding 3~\GeV\ and (b) reduced mass for preselected DVs with $\ntrk > 2$, in simulated signal samples
(dashed lines) and a background-enriched data sample (black points).  DVs in events satisfying the control region selection, defined in Section~\ref{sec:select}, are considered.  An arrow indicates the selection requirement which is applied to
the observable. The distributions are normalized to unit area. }
\label{fig:ntrk_mass}
\end{figure}

Preselected DVs are dominated by the decay products of inelastic interactions
between promptly produced particles and the detector material, and the decays of
SM particles with significant lifetime.
As shown in Figure~\ref{fig:ntrk_mass}, most of these DVs are rejected by
requiring that the number of tracks associated with the vertex $\ntrk$ is greater than
two.
The remaining DVs result primarily from the random crossings of unrelated
tracks in the vicinity of a metastable hadron decay.
Particularly in the case of large crossing angles, these random crossings
lead to artificially large values of the reconstructed vertex invariant mass.
To identify such vertices, tracks associated with DVs are removed one-by-one
and the angular separation $\Delta R$ between the momentum of the removed track and the
combined momentum of the remaining vertex tracks is calculated, with the observable \drmax defined as the
maximum of all such values.
The \textit{reduced mass} of the vertex is then calculated from the
ratio of the reconstructed vertex invariant mass, $m$, and \drmax.
As shown in Figure~\ref{fig:ntrk_mass}, the reduced mass discriminates strongly
between signal and background DVs for all values of \ma probed.
To reject vertices with random track crossings, DVs are required to have
$m/\drmax > 3$~\GeV.
 
To reduce the background contribution from heavy-flavour
hadron decays, additional cuts are placed on the significance of the vertex
radial position measurement, defined as the ratio of the vertex radial position
to its uncertainty $\sigma(r)$, and at least one track in
the vertex must have $|d_0| > 3$~mm with respect to the PV.
Finally, to facilitate the modelling of the background, a one-to-one matching between jets and DVs is performed, requiring DVs to be within $\Delta R = 0.6$ of one of the
four leading jets.
The full set of vertex selection criteria is shown in Table~\ref{table:vtx}.
 
\begin{table}[htb!]
\centering
\caption{The full set of selection criteria applied to DVs.}
\label{table:vtx}
\vspace{0.2cm}
\begin{tabular}{c | l } \hline \hline
Selection type & Requirement \\ \hline
Track pruning               & $|d^{\dv}_0| < 0.8$~mm\\
& $|z^{\dv}_0| < 1.2$~mm\\
& $\sigma(d^{\dv}_{0}) < 0.1$~mm\\
& $\sigma(z^{\dv}_{0}) < 0.2$ ~mm\\
\hline
Vertex preselection & $\chi^2/n_{\text{DoF}} < 5$\\
& $r < 300$~mm \\
& $|z| < 300$~mm\\
& pass material veto \\
\hline
Vertex selection & $n_{\text{trk}} > 2$ \\
& $m/\Delta R_{\text{max}}> 3$~\GeV\\
& $r/\sigma(r) > 100$  \\
& $\text{max}(|d_0|) > 3$~mm  \\
& $\Delta R_{\text{jet}} < 0.6$  \\
\hline
\hline
\end{tabular}
\end{table}


\section{Event selection}
\label{sec:select}

Events consistent with the $ZH,\ H \rightarrow aa$ topology were first selected
using the lowest-threshold unprescaled single-lepton ($e$, $\mu$)
triggers operating during each data-taking year~\cite{TRIG-2018-05,TRIG-2018-01}.
Events are then identified for reprocessing with the LRT and displaced vertex
reconstruction algorithms using an additional selection, referred to as a
\emph{filter}, which makes use of the results of the standard ATLAS
reconstruction algorithms to identify jets consistent with originating from an
LLP decay.
The selection requires the presence of at least one jet with $p_{\text{T}}>20$~\GeV, where
at this stage the full jet calibration is not yet applied and the jet energy is calibrated at the electromagnetic scale~\cite{ATLAS-CONF-2015-002}.
At least one of the two leading, central ($|\eta|<2.1$)
jets must be displaced.
Finally, the filter requires the presence of a lepton from the $Z$ boson decay.
The number of events passing this filter is sufficiently small to permit reconstruction
with the computationally expensive algorithms for displaced tracks and vertices.
A second filter is also used, which selects events passing a high-$p_{\text{T}}$
photon trigger, to facilitate the validation of the background estimation
method.
 
After reconstruction with the LRT algorithm, the signal and control regions are defined by
requiring events to contain two opposite-sign, same-flavour leptons.
The leading (sub-leading) lepton is required to have $p_{\text{T}} > 27$~(10)~\GeV.
To select events with a reconstructed $Z$ boson, the dilepton invariant mass
$m_{\ell\ell}$ is required to be in the range $66 < m_{\ell\ell} < 116$~\GeV.
Additionally, events are required to contain at least two jets.
The trigger, filter, lepton, $m_{\ell\ell}$, and jet requirements comprise the
event preselection.
The signal region (SR) is then defined by requiring that events pass the
preselection and contain at least two DVs passing the selections described in
Section~\ref{sec:sel_vtx}.
To ensure robust modelling of the background, the DVs are required to be
associated with different jets.
Depending on the value of $m_a$, these criteria select $ZH$ signal events with a leptonically-decaying $Z$ boson with an
efficiency ranging from $0.4\%$ to $0.6\%$ ($0.04\%$ to $0.2\%$) for $c\tau_a =10$~(100)~mm.
A control region (CR) is defined similarly to the SR except for inversion of
the DV multiplicity requirement, and is used to derive the background estimate.  Despite the relatively low efficiency for signal events to satisfy the SR selection, signal contamination in the CR is negligible.
 
The background estimate is validated using an orthogonal $\gamma+$jet event
selection.
This \emph{validation region} (VR) requires events to pass a high-$p_{\text{T}}$
photon trigger, contain at least one photon with $p_{\text{T}} >160$~\GeV\ or
two photons with $p_{\text{T}} >60$~\GeV, and at least two jets.
Additionally, events containing charged leptons are vetoed in the VR to remove
potential biases from signal.
No requirement is placed on the number of displaced jets in the VR.


\section{Background estimation}
\label{sec:background}

\begin{figure}[h!]
\centering
\includegraphics[width=0.7\textwidth]{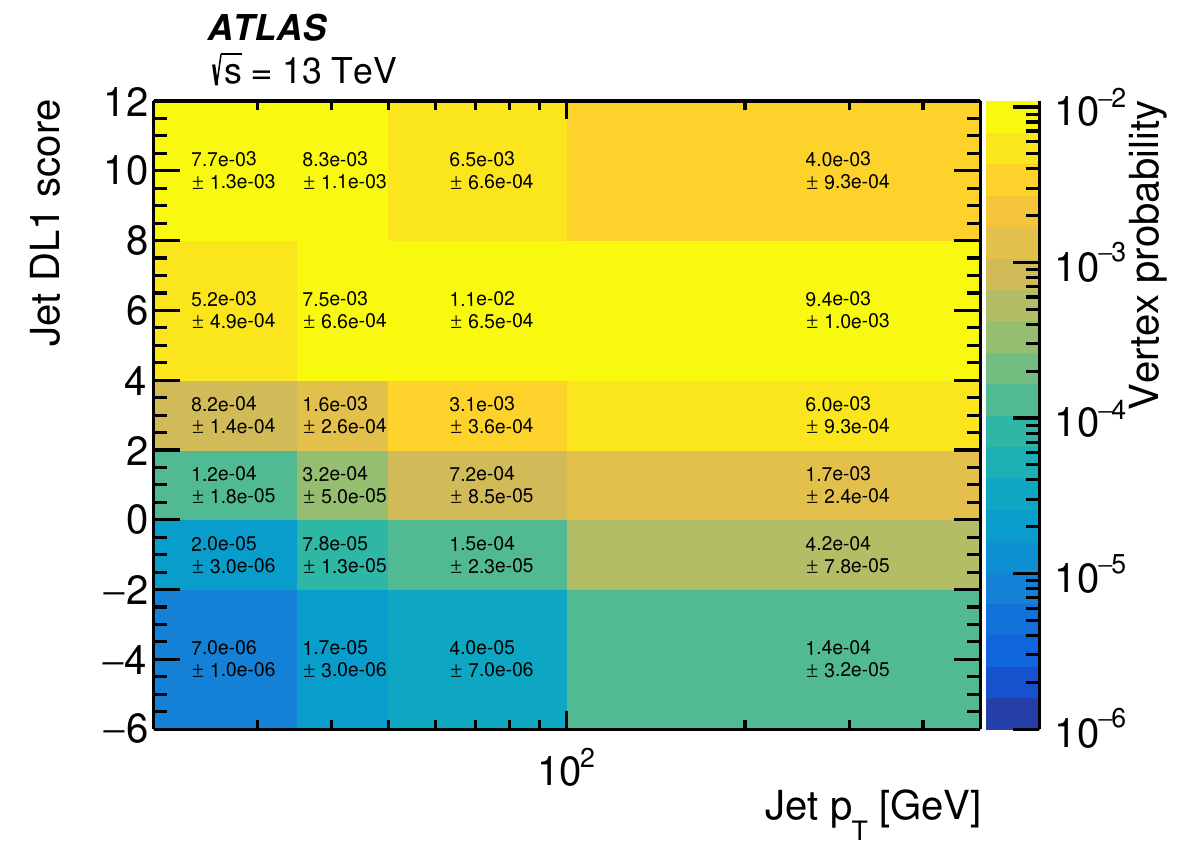}
\caption{The per-jet DV probabilities computed in the CR, as a function of jet $p_{\text{T}}$ and
DL1 $b$-tag score.  The rightmost column of bins is inclusive and includes jets with $p_{\text{T}} > 500$~\GeV.}
\label{fig:efficiencyMap}
\end{figure}

The probability for a jet to be matched to a DV within $\Delta R=0.6$ is
measured in the CR by taking the ratio of the number of jets matched to a DV to
the total number of jets.
This per-jet DV probability is observed to have a non-trivial dependence on the
properties of the jet.
The track multiplicity and density within a jet increase with the jet
$p_{\text{T}}$~\cite{STDM-2015-12}, causing this probability to increase correspondingly.
Due to the lifetime of hadrons containing heavy-flavour quarks, the probability
also depends strongly on the flavour of the parton that initiated the jet.
Therefore, the per-jet DV probability $P_{\text{DV}}(j)$ is
parameterized as a function of jet $p_{\text{T}}$ and DL1 $b$-tagging discriminant, as
shown in Figure~\ref{fig:efficiencyMap}.
 
By restricting the computation of the per-jet DV probability to a region requiring fewer than
two DVs, a bias of the order of $P_{\text{DV}}^2$ is introduced.
However, due to the fact that $P_{\text{DV}} \ll 1$, these higher-order
corrections are negligible.
 
The probability that an event contains a given number of DVs among the four
leading jets may then be computed from a multinomial distribution.
For example, the probability that an event contains exactly one DV is given by:
\begin{equation*}
\label{eq:prob}
P_{1} = \sum_{i = 1}^4 P_{\text{DV}}(j_i) \times \prod_{k \neq i} \left(1 - P_{\text{DV}}(j_k) \right).
\end{equation*}
If the event contains fewer than four jets, then the probability is computed
from all jets in the event.
The probability for an event to fall in the SR is given by $P_{\geq 2} = 1 - P_{0} - P_{1}$.
Therefore, the background estimate is obtained by weighting preselected data events by
$P_{\geq 2}$.
 
The presence of signal contamination in preselected data would increase the per-jet DV probability, as well as the population of events to which this probability is applied.  Both of these effects would artificially increase the background estimate, potentially concealing the presence of signal in the SR.  This effect was studied with a conservative assumption about the branching ratio $\mathcal{B}(H \rightarrow aa)$ and found to have less than a 10\% effect on the background prediction, negligible in comparison to the size of injected signal.
 
To compute the uncertainty in the background estimate, pseudo-experiments are
performed.
Each pseudo-experiement is assigned a unique set of per-jet probabilities by
varying each $(p_{\text{T}},\mathrm{DL1})$ bin within statistical
uncertainties, and the varied probabilities are used to predict the background
yield.
The mean of the pseudo-experiments determines the central value of the expected
yield, with its statistical uncertainty given by the RMS of pseudo-experiments.
Using this method, a background estimate of  $1.30 \pm 0.08$ events is
obtained, where the uncertainty is statistical only.
 
This method for computing the background prediction relies on the assumption
that the per-jet DV probabilities are unaffected by the presence of multiple
DVs in an event.
To validate this assumption, the VR defined in Section~\ref{sec:select}
is used.
The parameterization of the per-jet DV probabilities in terms of $p_{\text{T}}$
and DL1 allows for a robust validation of the method despite the kinematic
properties of the jets in this region being different from those in the SR, and
the absence of the displaced jet requirement leaves a statistically
significant number of events to study.
By comparing the predicted and observed numbers of events with at least two
DVs in the VR, the core assumption of the background estimate may be tested
and any systematic differences may be taken into account as an additional
uncertainty in the background estimate.
 
\begin{figure}
\centering
\includegraphics[width=0.6\textwidth]{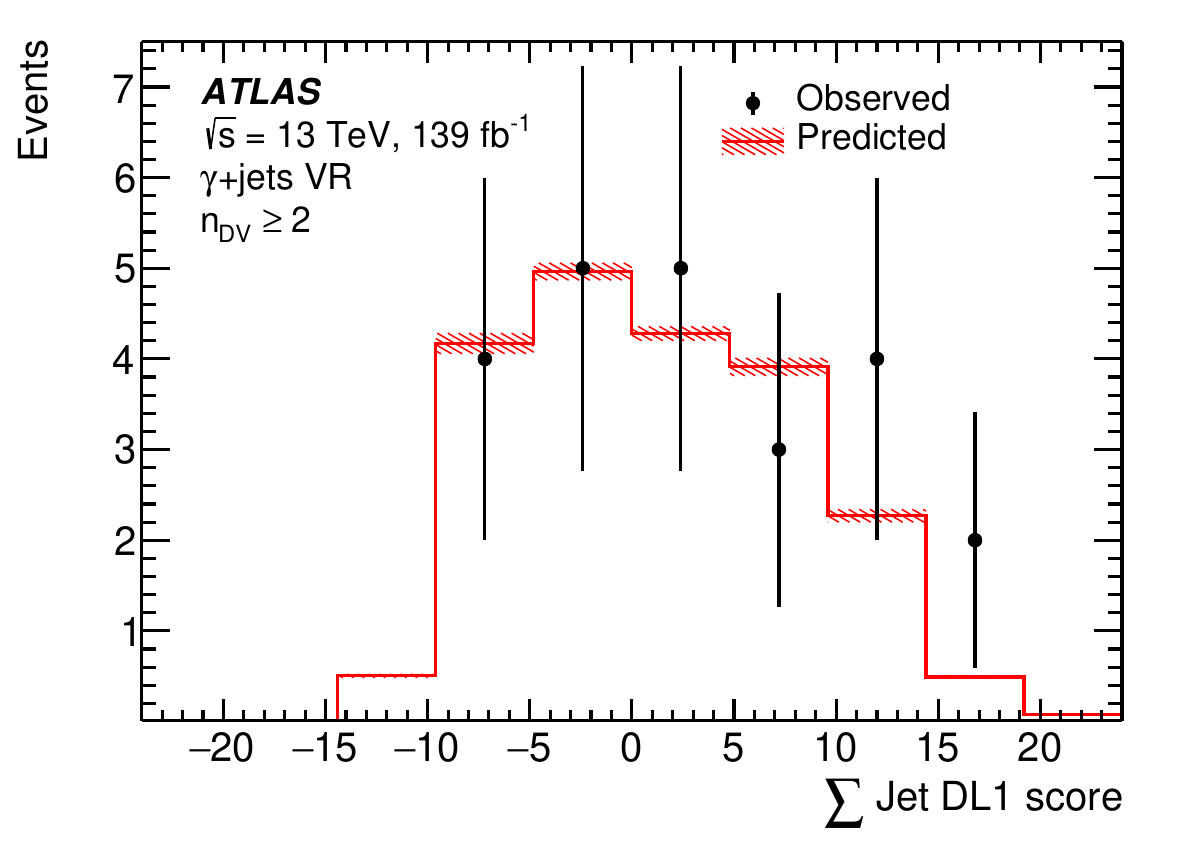}
\caption{The predicted (solid line) and observed (black points) numbers of background events with at least
two DVs in the $\gamma+$jet VR as a function of the sum of the DL1 $b$-tag scores of the (up to) four leading jets. The shaded band represents the statistical uncertainty of the prediction.}
\label{fig:gamma}
\end{figure}

Following the same procedure as is used in the CR, the per-jet DV probability
is calculated using events in the VR with fewer than two DVs, and used to
predict the number of events with at least two DVs.
Using this method, a mean of $19.9 \pm 0.4$ events are predicted and 23 are observed, as
shown in Figure~\ref{fig:gamma}, demonstrating that any jet--jet correlations
are negligible or captured by the parameterization in terms of jet $p_{\text{T}}$ and
DL1.
 
The systematic uncertainty of the method used to obtain the background prediction is
derived by comparing the numbers of predicted and observed events in the VR.
Although the prediction and observation agree within uncertainties, not many events
are available to test the background estimation method. Consequently, a
conservative systematic uncertainty, equal to the statistical uncertainty of
the observed number of events in the VR, is assigned to the total expected number of
background events in the SR.
This amounts to a 21\% uncertainty, or $\pm 0.27$ events.
This 21\% uncertainty also covers the sensitivity of the background estimate to
possible signal contamination impacting the per-jet DV probabilities, as discussed above.
 
Additional studies were performed by considering the subset of VR events that contain a displaced jet.
This requirement reduces the statistical power of the VR, but makes it more
representative of the SR.
With the addition of this requirement, 2 events are observed compared to a
prediction of $2.20 \pm 0.10 \, \mathrm{(stat.)}$, and the per-jet DV probabilities are
compatible with those derived in the control region.
This study provides further validation that the core assumptions of the
background estimation method are robust against different event and
object-level selections, but is ultimately not considered when deriving the
systematic uncertainty on the background estimate due to the limited
available statistics.


\section{Signal systematic uncertainties}
\label{sec:syst}

 
While the analysis is dominated by the statistical uncertainty of the observed data yield, several sources of experimental and theoretical uncertainty are considered that could potentially affect the signal efficiency.
 
Experimental uncertainties originating from differences between data and
simulation are considered for all physics objects used in the analysis.
For standard objects, these include uncertainties in the reconstruction and
identification efficiencies, as well as the energy calibrations.
Efficiency scale factors are used to correct the modelling of electrons and muons in
simulation relative to data.
These comprise electron identification and muon reconstruction scale factors as
well as the scale factors applied to correct for trigger efficiency differences between data
and simulation~\cite{EGAM-2018-01,PERF-2015-10,TRIG-2018-05,TRIG-2018-01}.
The associated uncertainties are computed by varying the scale factors up and down
by one standard deviation and evaluating the impact on the final expected
number of signal events.
For jets, jet energy scale (JES) and jet energy resolution (JER) uncertainties
are considered~\cite{JETM-2018-05}.
The applicability of the standard jet calibration scheme to the displaced jets
considered in this analysis has been studied and found to be
satisfactory~\cite{ATL-PHYS-PUB-2019-025}.
Simulated events are reweighted such that the distribution of the average
number of interactions per bunch crossing matches the distribution measured in
data.
Since the number of interactions per bunch crossing depends on the inelastic $pp$
cross section, the difference between the predicted and measured cross sections~\cite{STDM-2015-05}
is propagated to the simulation by a systematic variation of the reweighted distribution.
 
The dominant systematic uncertainty in the expected signal yield originates
from the non-standard reconstruction methods used.
The uncertainty in the reconstruction of displaced vertices comes from
the difference in performance of both the standard and LRT
algorithms between data and simulation.
For standard tracking, this uncertainty is known to be around 2\% for tracks
with small displacements from the IP~\cite{ATL-PHYS-PUB-2015-051}.
To assess the uncertainty in the vertex reconstruction efficiency due to the
modeling in the LRT algorithm, the rates of displaced vertices consistent with
$K^0_{\mathrm{S}} \rightarrow \pi^+ \pi^-$ decays are compared between data and
$Z$+jets simulation.
The uncertainty is estimated by examining the difference in $K^0_{\mathrm{S}}$ yield
between data and simulation.
From the preselected events, candidate $K^0_{\mathrm{S}}$ vertices are
identified by requiring that the vertices pass the vertex preselection, have
exactly two tracks of opposite charge, and have an invariant mass in the region
450 to 550~\MeV.
The simulation is normalized such that the number of $K^0_{\mathrm{S}}$ vertices inside
the beam pipe ($r < 23.5$~mm) is the same in data and simulation.
This accounts for any $K^0_{\mathrm{S}}$ decay yield differences that may exist
between data and simulation in a region where tracking uncertainties are well understood.
The $p_{\text{T}}$ distributions of tracks from the $K^0_{\mathrm{S}}$ vertices were found
to be consistent between data and simulation.
The vertex yields of $K^0_{\mathrm{S}}$ are then compared between data and the
normalized simulation in five bins of $r$ ranging from zero to the radius of the last
pixel layer (122~mm), beyond which there are significantly fewer vertices.
The ratio quantifies the discrepancy between data and simulation in the number of
$K^0_{\mathrm{S}}$ vertices reconstructed from tracks outside of the beam pipe.
To compute a per-track uncertainty, the square root of the ratio is taken because
$K^0_{\mathrm{S}}$ vertices are two-track vertices.
This uncertainty is then added in quadrature with the 2\% uncertainty from the
standard tracking pass.
 
To propagate this uncertainty to the signal yield, tracks are randomly removed
from reconstructed vertices in an uncorrelated fashion\footnote{A fully correlated method in which the entire vertex is removed was also investigated and gave similar results, but is believed
to overestimate the uncertainty.} according to a uniform probability given
by the per-track uncertainty corresponding to the $r$ position of the
vertex.
The tracks are removed following the same track-pruning procedure as described in
Section~\ref{sec:sel_vtx}, and the vertex properties are recalculated
from the updated set of associated tracks.
The yield of vertices passing the full signal vertex
selection is then compared between the modified and nominal vertex collections
as a function of $r$, and the ratio of yields provides a scale factor corresponding to
the reconstruction efficiency of each LLP.
A per-event efficiency correction scale factor is obtained from the product of the
scale factors corresponding to each LLP.
The difference between the efficiency computed with the scale factors applied
and the nominal selection efficiency is taken as the systematic uncertainty.
This uncertainty is largest for small values of $m_a$ and increases with the
proper lifetime $c\tau_a$ to a maximum value of 12\%.
 
Uncertainties associated with the displaced-jet filter are considered.
The dominant contribution to the uncertainty originates from the use of uncalibrated jets
in the filter.
This uncertainty is estimated by raising the jet $p_{\text{T}}$ threshold in the filter
by 5~\GeV, from 20 to 25~\GeV, and calculating the change in filter efficiency on
simulated signal samples.
The choice of 5~\GeV\ is motivated by studying the difference in efficiency of
the uncalibrated jet \pt selection between data and simulation as a function of calibrated jet \pt.
 
A theoretical uncertainty of $\pm 4\%$ and $\pm 25\%$ is assumed for the total
$q\bar{q} \rightarrow ZH$ and $gg \rightarrow ZH$, $H \rightarrow aa$
cross section, respectively~\cite{deFlorian:2016spz}.
These uncertainties include effects from varying the factorization and renormalization
scales, the PDF and \alphas.
To determine the uncertainty due to the choice of parton shower model, the nominal
\PYTHIA[8] samples are compared with a reference \HERWIG[7] sample of $ZH$ production.
The $p_{\text{T}}$ of the $ZH$ system is compared between the two samples and
used to reweight events in the \PYTHIA signal samples to the reference \HERWIG
sample.
The change in acceptance after reweighting is found to be of the order of
$0.5\%$ and is included as an additional systematic uncertainty in the signal
modelling.
 
Finally, the expected number of signal events is subject to the uncertainty in
the total integrated luminosity of the data sample used.
The uncertainty in the combined 2015--2018 integrated luminosity is 1.7\%
\cite{ATLAS-CONF-2019-021}, obtained using the LUCID-2 detector \cite{Avoni_2018}
for the primary luminosity measurements.
 
A summary of the systematic uncertainties is given in Table~\ref{table:syst}.
With the exception of the uncertainties from the LRT algorithm and the displaced jet filter, the
systematic uncertainties are not observed to vary significantly with $m_{a}$
or $c\tau_{a}$ and are thus computed from the weighted average across samples.
\begin{table}[h]
\centering
\caption{Summary of all sources of systematic uncertainty considered in the analysis that affect the signal yield.
With the exception of the uncertainties from the filter and the LRT algorithm,
no significant dependence on $m_a$ or $c\tau_{a}$ is observed, so the quoted
values are those derived from averaging over the different masses and
lifetimes.  The uncertainty from the filter efficiency was found to be
uncorrelated with $c\tau_{a}$ but to increase with $m_{a}$, so the range quoted corresponds to the observed variation as $m_a$ increases, after averaging over $c\tau_{a}$.
The uncertainty from the LRT algorithm was found to be correlated with both $c\tau_{a}$ and $m_a$, so the quoted range corresponds to the observed variation with $m_a$ and $c\tau_{a}$ (with the uncertainty increasing most strongly with $c\tau_{a}$).}
\label{table:syst}
\vspace{0.2cm}
\small
\begin{tabular}{ l | c }
\hline
\hline
Source & Uncertainty [\%] \\
\hline
Theory                  & 4.7 \\
\hline
Luminosity              & 1.7 \\
\hline
Pile-up reweighting      & 2.6 \\
\hline
Electron identification & 1.6 \\
Electron calibration    & 0.4 \\
Muon reconstruction     & 0.8 \\
Muon calibration        & 0.4 \\
\hline
Electron trigger        & 0.7 \\
Muon trigger            & 1.3 \\
\hline
Jet energy scale        & 1.4 \\
Jet energy resolution   & 1.3 \\
\hline
Filter                  & 2.8--3.8 \\
\hline
LRT                     & 2.4--12~  \\
\hline
Total                   & 7.4--14~  \\
\hline
\hline
\end{tabular}
\end{table}


\section{Results}
\label{sec:result}

\begin{figure}
\centering
\includegraphics[width=0.6\textwidth]{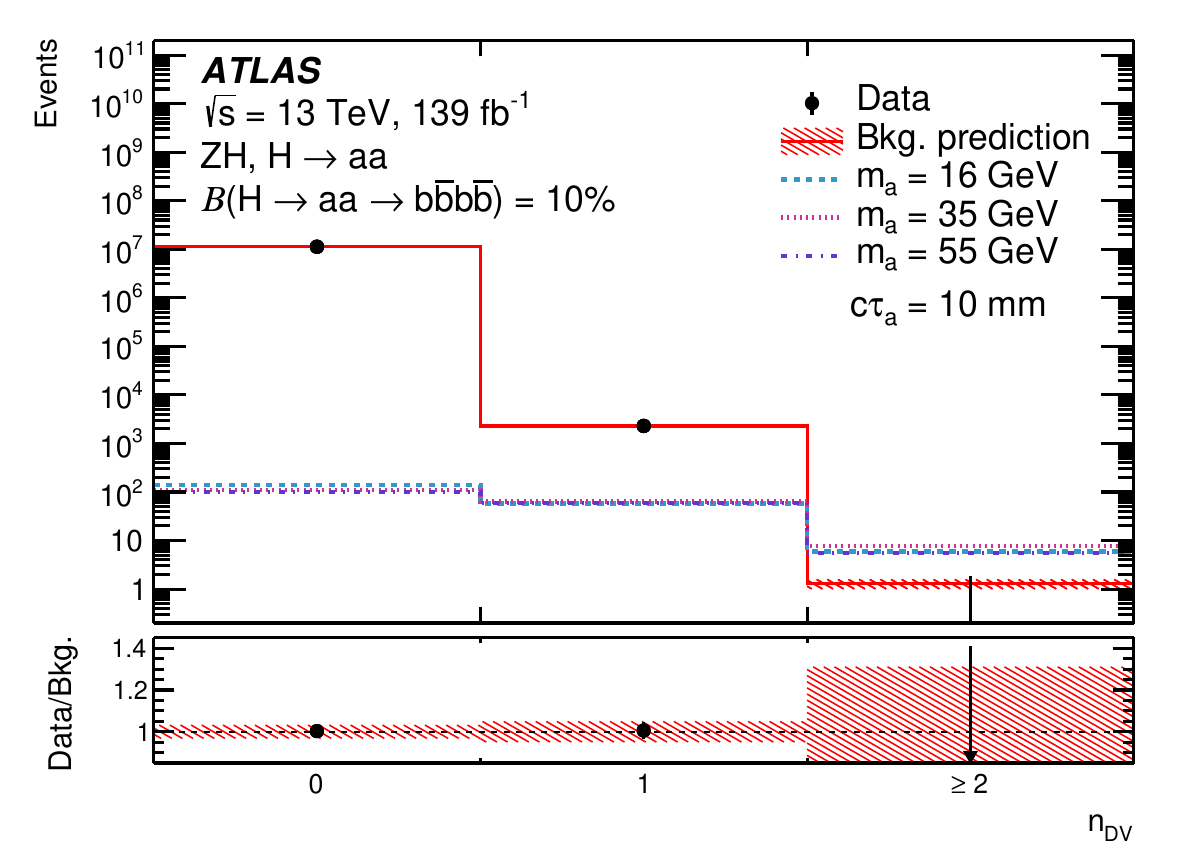}
\caption{DV multiplicity among preselected events for signal (dashed lines),
background prediction (solid line), and data (black points).  The bins
corresponding to $n_{\text{DV}} = 0$ and $n_{\text{DV}} = 1$ compose the CR, which is used to
derive the background estimate. The third bin is the SR and contains all
events with $n_{\text{DV}} \geq 2$.  The shaded bands represent the combined
statistical and systematic uncertainty of the prediction.  Signal
distributions are normalized assuming $\mathcal{B}(H\rightarrow aa
\rightarrow b\bar{b}b\bar{b})=10\%$.}
\label{fig:nDV}
\end{figure}
 
The mean number of background events predicted in the SR is $1.30 \pm 0.08\text{ (stat.)}\pm 0.27 \text{ (syst.)}$.
Zero events are observed in the signal region, compatible with the background prediction at the 68\% confidence level (CL), as shown in Figure~\ref{fig:nDV}.
 
Upper limits at the 95\% CL are set on the branching ratio of
the Higgs boson to pairs of long-lived pseudoscalars $\mathcal{B}(H\rightarrow
aa \rightarrow b\bar{b}b\bar{b})$ for each signal mass hypotheses,
following the $\text{CL}_\text{s}$ prescription~\cite{Read:2002hq} with a
profile likelihood-ratio test statistic.
A Poisson probability term describing the total number of observed events is
used, with the systematic uncertainties of the signal and background yields
treated as nuisance parameters and assigned Gaussian constraints with the
appropriate widths, as described in Section~\ref{sec:syst}.
Pseudo-experiments which sample the distribution of the profile
likelihood ratio are generated to compute the $p$-value and derive the exclusion limits.
\begin{figure}
\centering
\includegraphics[width=0.6\textwidth]{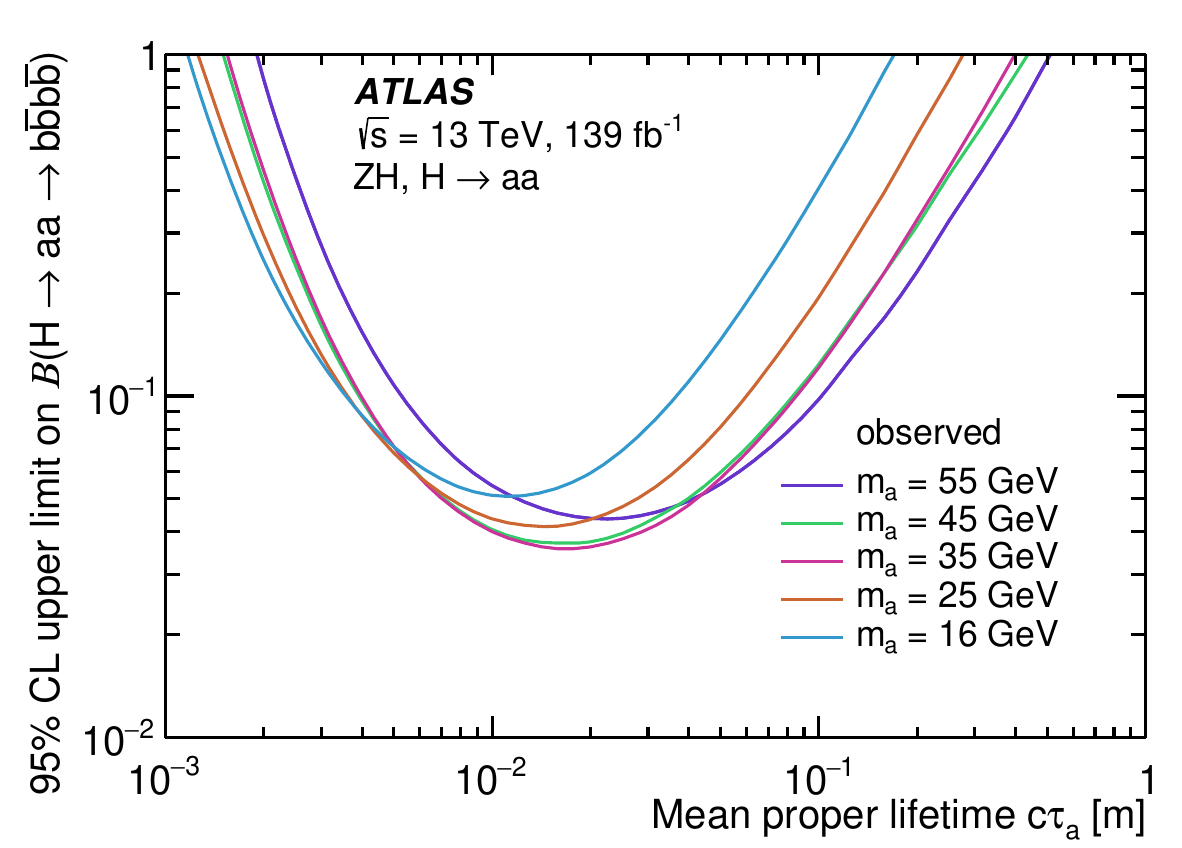}
\caption{Observed $95\%$ CL exclusion limits on $\mathcal{B}(H\rightarrow aa
\rightarrow b\bar{b}b\bar{b})$ as a function of $c\tau_a$.}
\label{fig:limit}
\end{figure}
The observed limits for all signal mass points as a function of $c\tau_a$ are
shown in Figure~\ref{fig:limit}.
 
The limits are strongest for $\ctaua\approx$ 10--20~mm, probing values of $\mathcal{B}(H\rightarrow aa\rightarrow b\bar{b}b\bar{b})$ below $\sim\!4\%$, and become weaker for shorter/longer lifetimes, primarily due to reduced radial acceptance.  The limits depend only weakly on the $a$ boson mass, due to the relative insensitivity of the reduced-mass observable to $m_a$.
 
For $m_a=16$~(55)~\GeV, a branching ratio greater than 10\% is excluded at the 95\% CL for $3.7 < c\tau_a < 37$~mm ($5.4 < c\tau_a< 102$~mm), probing an important gap in the ATLAS exotic Higgs decay programme. Decays of the Higgs boson into LLPs with shorter lifetimes are probed by conventional searches using standard $b$-tagging algorithms to identify signal events~\cite{HIGG-2017-05,HDBS-2018-47}, while decays into LLPs with longer lifetimes are probed by calorimeter- and MS-based searches utilizing dedicated LLP triggers~\cite{EXOT-2017-25,EXOT-2017-05,EXOT-2018-61}.


\FloatBarrier
 
\section{Conclusion}
\label{sec:conclusion}

A search for exotic decays of the Higgs boson into pairs of long-lived
pseudoscalars $(a)$ is presented, using $139~\text{ fb}^{-1}$ of $\sqrt{s} =
13$~\TeV\ $pp$ collision data collected between 2015 and 2018 with the ATLAS
experiment at the LHC.
The search is performed in the topology of associated $ZH$ production.
Displaced vertices with high mass and track multiplicity relative to ones from
SM processes are exploited as a signature of $a$~boson decays and
as a means to reject backgrounds.
A data-driven method is used to assess the expected number of background events
satisfying the event selection.
The number of observed events is consistent with this prediction, and limits on the branching ratio
$\mathcal{B}(H\rightarrow aa \rightarrow b\bar{b}b\bar{b})$ are calculated at the 95\% CL.
For masses below $40$~\GeV, these are the most stringent limits to date in the probed lifetime regime.


\section*{Acknowledgements}


We thank CERN for the very successful operation of the LHC, as well as the
support staff from our institutions without whom ATLAS could not be
operated efficiently.
 
We acknowledge the support of
ANPCyT, Argentina;
YerPhI, Armenia;
ARC, Australia;
BMWFW and FWF, Austria;
ANAS, Azerbaijan;
SSTC, Belarus;
CNPq and FAPESP, Brazil;
NSERC, NRC and CFI, Canada;
CERN;
ANID, Chile;
CAS, MOST and NSFC, China;
Minciencias, Colombia;
MSMT CR, MPO CR and VSC CR, Czech Republic;
DNRF and DNSRC, Denmark;
IN2P3-CNRS and CEA-DRF/IRFU, France;
SRNSFG, Georgia;
BMBF, HGF and MPG, Germany;
GSRI, Greece;
RGC and Hong Kong SAR, China;
ISF and Benoziyo Center, Israel;
INFN, Italy;
MEXT and JSPS, Japan;
CNRST, Morocco;
NWO, Netherlands;
RCN, Norway;
MEiN, Poland;
FCT, Portugal;
MNE/IFA, Romania;
JINR;
MES of Russia and NRC KI, Russian Federation;
MESTD, Serbia;
MSSR, Slovakia;
ARRS and MIZ\v{S}, Slovenia;
DSI/NRF, South Africa;
MICINN, Spain;
SRC and Wallenberg Foundation, Sweden;
SERI, SNSF and Cantons of Bern and Geneva, Switzerland;
MOST, Taiwan;
TAEK, Turkey;
STFC, United Kingdom;
DOE and NSF, United States of America.
In addition, individual groups and members have received support from
BCKDF, CANARIE, Compute Canada and CRC, Canada;
COST, ERC, ERDF, Horizon 2020 and Marie Sk{\l}odowska-Curie Actions, European Union;
Investissements d'Avenir Labex, Investissements d'Avenir Idex and ANR, France;
DFG and AvH Foundation, Germany;
Herakleitos, Thales and Aristeia programmes co-financed by EU-ESF and the Greek NSRF, Greece;
BSF-NSF and GIF, Israel;
Norwegian Financial Mechanism 2014-2021, Norway;
NCN and NAWA, Poland;
La Caixa Banking Foundation, CERCA Programme Generalitat de Catalunya and PROMETEO and GenT Programmes Generalitat Valenciana, Spain;
G\"{o}ran Gustafssons Stiftelse, Sweden;
The Royal Society and Leverhulme Trust, United Kingdom.
 
The crucial computing support from all WLCG partners is acknowledged gratefully, in particular from CERN, the ATLAS Tier-1 facilities at TRIUMF (Canada), NDGF (Denmark, Norway, Sweden), CC-IN2P3 (France), KIT/GridKA (Germany), INFN-CNAF (Italy), NL-T1 (Netherlands), PIC (Spain), ASGC (Taiwan), RAL (UK) and BNL (USA), the Tier-2 facilities worldwide and large non-WLCG resource providers. Major contributors of computing resources are listed in Ref.~\cite{ATL-SOFT-PUB-2021-003}.


\printbibliography
 
\clearpage 
 
\begin{flushleft}
\hypersetup{urlcolor=black}
{\Large The ATLAS Collaboration}

\bigskip

\AtlasOrcid[0000-0002-6665-4934]{G.~Aad}$^\textrm{\scriptsize 100}$,
\AtlasOrcid[0000-0002-5888-2734]{B.~Abbott}$^\textrm{\scriptsize 117}$,
\AtlasOrcid[0000-0002-7248-3203]{D.C.~Abbott}$^\textrm{\scriptsize 101}$,
\AtlasOrcid[0000-0002-2788-3822]{A.~Abed~Abud}$^\textrm{\scriptsize 34}$,
\AtlasOrcid[0000-0002-1002-1652]{K.~Abeling}$^\textrm{\scriptsize 53}$,
\AtlasOrcid[0000-0002-2987-4006]{D.K.~Abhayasinghe}$^\textrm{\scriptsize 92}$,
\AtlasOrcid[0000-0002-8496-9294]{S.H.~Abidi}$^\textrm{\scriptsize 27}$,
\AtlasOrcid[0000-0001-5329-6640]{H.~Abramowicz}$^\textrm{\scriptsize 148}$,
\AtlasOrcid[0000-0002-1599-2896]{H.~Abreu}$^\textrm{\scriptsize 147}$,
\AtlasOrcid[0000-0003-0403-3697]{Y.~Abulaiti}$^\textrm{\scriptsize 5}$
% ...
\bigskip
\\

% ...
$^{5}$High Energy Physics Division, Argonne National Laboratory, Argonne IL; United States of America.\\
% ...
$^{27}$Physics Department, Brookhaven National Laboratory, Upton NY; United States of America.\\
% ...
$^{34}$CERN, Geneva; Switzerland.\\
% ...
$^{53}$II. Physikalisches Institut, Georg-August-Universit\"{a}t G\"ottingen, G\"ottingen; Germany.\\
% ...
$^{92}$Department of Physics, Royal Holloway University of London, Egham; United Kingdom.\\
% ...
$^{100}$CPPM, Aix-Marseille Universit\'e, CNRS/IN2P3, Marseille; France.\\
$^{101}$Department of Physics, University of Massachusetts, Amherst MA; United States of America.\\
% ...
$^{117}$Homer L. Dodge Department of Physics and Astronomy, University of Oklahoma, Norman OK; United States of America.\\
% ...
$^{147}$Department of Physics, Technion, Israel Institute of Technology, Haifa; Israel.\\
$^{148}$Raymond and Beverly Sackler School of Physics and Astronomy, Tel Aviv University, Tel Aviv; Israel.\\
% ...

\end{flushleft}


\end{document}